\documentclass[aps,prb,floatfix,amsmath,amssymb,superscriptaddress,twocolumn,nofootinbib]{revtex4-2}
\usepackage{bm}                     
\usepackage{appendix}
\usepackage[latin1]{inputenc}
\usepackage{dcolumn}      
\usepackage{bbm}
\usepackage{color}
\usepackage{xcolor}

\usepackage[mathscr]{eucal}
\usepackage{latexsym}
\usepackage{amsbsy}
\usepackage{float}
\usepackage{graphicx}
\usepackage{epsfig}
\usepackage{subfigure}
\usepackage{wrapfig}
\usepackage{epsf}
\usepackage{float}
\usepackage[normalem]{ulem}
\usepackage{qcircuit}

\definecolor{darkblue}{HTML}{004D6B}
\definecolor{darkred}{HTML}{8c1515}
\definecolor{darkgreen}{HTML}{006400}
\usepackage{hyperref}
\hypersetup{ colorlinks=true, urlcolor=darkblue, citecolor=darkred,
    linkcolor=darkblue, breaklinks }
\usepackage{parskip}

\newcommand{\be}{\begin{equation}}
\newcommand{\ee}{\end{equation}}
\newcommand{\ba}{\begin{array}{l}}
\newcommand{\ea}{\end{array}}
\newcommand{\re}[1]{(\ref{#1})}

\newcommand{\banonum}{\begin{eqnarray*}}
\newcommand{\eanonum}{\end{eqnarray*}}
\newcommand{\baa}{\begin{eqnarray}}
\newcommand{\eaa}{\end{eqnarray}}
\newcommand{\bfr}{\begin{flushright}}
\newcommand{\efr}{\end{flushright}}
\newcommand{\bfl}{\begin{flushleft}}
\newcommand{\efl}{\end{flushleft}}
\newcommand{\lab}[1]{\label{#1}}

\begin{document}
\title{Coulomb impurities in graphene driven by fast ions}

\author{Saparboy Rakhmanov}
\affiliation{Chirchik State Pedagogical University, 104 Amur Temur Str., 111700 Chirchik, Uzbekistan}
\author{Reinhold Egger}
\affiliation{Institut f\"ur Theoretische Physik, Heinrich-Heine-Universit\"at, 40225  D\"usseldorf, Germany}
\author{Doniyor Jumanazarov}
\affiliation{Department of Physics, Urgench State University, Urgench 220100, Uzbekistan}
\affiliation{Department of Telecommunication Engineering, Urgench Branch of Tashkent University of Information Technologies, Urgench 220100, Uzbekistan}
\author{Davron Matrasulov}
\affiliation{Turin Polytechnic University in Tashkent, 17 Niyazov Str., 100095 Tashkent, Uzbekistan}
\affiliation{Center for Theoretical Physics, Khazar University, 41 Mehseti Street, Baku, AZ1096, Azerbaijan}

\begin{abstract}
We provide a theoretical model for electronic transitions in a two-dimensional (2D) artificial atom in a graphene monolayer. The artificial atom is due to the presence of a charged adatom (Coulomb impurity) in the layer and interacts 
with a fast ultrarelativistic ion moving parallel to the layer. We compute the probability and cross sections for the corresponding electronic transitions by means of an exact solution of the time-dependent 2D Dirac equation describing the interaction of the planar atom with the electromagnetic field of the ultrarelativistic projectile. 
\end{abstract}
\maketitle

\maketitle

It is well known that by doping charged impurities into graphene monolayers, one can create artificial planar relativistic atoms described by the Dirac equation  \cite{CastroNeto}. Coulomb impurities in graphene thus provide a  powerful testing ground for relativistic quantum mechanics and  (2+1) dimensional quantum electrodynamics (QED)  
 \cite{Luican,Wang,Wang2,Pereira,Gamayun,Dipole,Denis,Rakhmanov24}. In addition, high-energy phenomena such as 2D versions of the electronic transitions induced by relativistic ion-atom collisions can be studied for graphene at the table-top level. For relativistic ion-atom collisions, electronic transitions may appear in the form of excitation, ionization and/or electron-positron (\emph{aka} electron-hole) creation processes. In the conventional 3D case, such high-energy collision phenomena can be only probed in collider facilities such as the LHC (CERN) \cite{Renk11}, the Relativistic Heavy Ion Collider (Brookhaven) \cite{Abdulhamid24,Adam20}, or the FAIR facility at GSI \cite{Thomas2015,Hill2013,Hag2013}. 

\begin{figure}[b]
\centering
\includegraphics[width=0.5\textwidth]{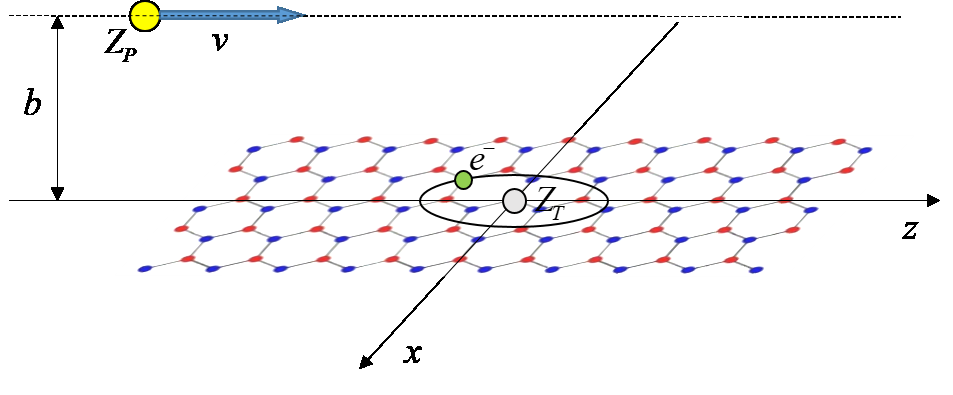}
 \caption{Schematic sketch of the geometry studied in this paper. A fast charged projectile, represented by an ion with charge $Z_P e$, moves parallel to the graphene layer at perpendicular distance $b$ and speed $v$. 
 The ion interacts with an artifical 2D atom realized by doping an adatom with charge $Z_Te$ (Coulomb impurity) into the layer. The graphene layer here corresponds to the $xz$ plane, with the artificial atom at the origin.  The ion trajectory is assumed to be at $x=0$, with the velocity pointing along the $z$ direction.}
 \label{fig1}
\end{figure}

In  this work, we consider an artificial atom due to a Coulomb impurity with charge $Z_T e$ in the graphene layer.
The atom interacts with a fast ion of charge $Z_P e$ moving parallel to 
the graphene layer at perpendicular distance $b$ and speed $v$, see Fig.~\ref{fig1} for an illustration.
We use coordinates where the $xz$ plane represents the graphene layer, the Coulomb impurity is centered at $x=z=0$, and the ion propagates along the $z$ direction with $x=0$ and $y=b$.
We study the electronic transitions induced in the artificial atom by the sudden perturbation of the time-dependent
Coulomb field generated by the ion. Our study has been inspired by a series of works by Baltz and co-authors \cite{Baltz00,Baltz01,Baltz03,Baltz05,Baltz91,Baltz93,Baltz95,Baltz97,Baltz98,Baltz93_02}, where the probabilities and cross-sections of different electronic transitions induced by the collision of a relativistic atom with a fast charged ion have been computed. Considering the ultrarelativistic limit of the projectile's velocity, Baltz \emph{et al.}~have obtained an exact solution of the time-dependent Dirac equation describing the interaction of a fast ion with the relativistic atom.  We here consider the 2D counterpart of their analysis.

\emph{Relativistic 2D atom in graphene driven by a fast ion.---}Coulomb impurities in graphene  have attracted much attention as condensed matter based realizations of relativistic quantum physics and $(2+1)$ dimensional QED. 
Basic mechanisms for generating vacuum polarization effects (such as quasiparticle pair creation) induced by time-dependent electromagnetic fields have been studied theoretically in Refs.~\cite{Kotov,Novikov,Khalilov,Allor,Lewk,Lewk1,Most1,Fillion,Akal01,Akal02,Golub,selym}.
In particular, the critical charge for supercriticality, at which the lowest bound-state energy level dives
into the filled Dirac sea, turns out to be much lower than for 3D atoms \cite{Novikov,Khalilov}.
In the absence of the ion beam, the artificial 2D atom in graphene has been described in terms of the  
 Dirac equation by Novikov \cite{Novikov}, where  both the discrete spectrum and the scattering states have been calculated.  Following Ref.~\cite{Novikov}, we first briefly recall the Dirac equation for the electronic motion in graphene in the presence of a Coulomb impurity. For the vacuum counterpart of this system, see Ref.~\cite{Khalilov}. 
Allowing for a homogeneous quasiparticle energy gap $M$, which could be generated by strain effects due to an underlying
substrate, the low-energy Hamiltonian is given by \cite{CastroNeto,Novikov,Kotov}
\be
H_0 = v_F(\sigma_x p_x +\sigma_z p_z)+M\sigma_z -\frac{v_F\alpha_{gr} Z_T}{r} \sigma_0 .
\lab{hamilt}
\ee 
Here $v_F\approx c/300$ denotes the Fermi velocity in ungapped graphene ($c$ is the velocity of light),  $Z_T e$ is the charge of the Coulomb impurity (target), and $\alpha_{gr}\sim {\cal O}(1)$ is the effective fine structure constant of the graphene layer which depends on the dielectric constant of the substrate. The Pauli matrices $\sigma_{x,z}$ (with identity $\sigma_0$) act in sublattice space for graphene's honeycomb lattice \cite{CastroNeto}. We use ${\bf r}=(x,z)$, $r=\sqrt{x^2+z^2}$, and $p_{x,z} =-i\partial_{x,z}$.  
For clarity, we consider a single $K$ valley of the band structure and a single spin polarization. These
assumptions are justified if the electromagnetic field generated by the ion pulse varies smoothly in space
on the scale of the lattice constant ($a_0\sim 2.46$~\AA) \cite{CastroNeto}.

In the absence of external fields, the quasiparticle 
spinor wave function obeys the stationary Dirac equation, $H_0|\psi\rangle =E|\psi\rangle$.
Following Refs.~\cite{Novikov,Khalilov}, we separate angular ($\phi$) and radial ($r$) variables,
\be \label{FG}
\psi_{n,j}({\bf r})=\left(\begin{array}{c}F_{n,j}(r) \,\Phi_{j-1/2}(\phi)\\
iG_{n,j}(r)\, \Phi_{j+1/2}(\phi)\end{array}\right),
\ee
where the integers $n$ and half-integers $j$ are the principal and angular momentum quantum numbers, respectively.
With integer $m$, the angular eigenfunctions are $\Phi_m(\phi)=\frac{1}{\sqrt{2\pi}}e^{im\phi}$. 
The radial eigenfunctions $F_{n,j}(r)$ and $G_{n,j}(r)$ are specified in closed form in Ref.~\cite{Novikov}.
We here focus on the discrete spectrum, $|E|<M$. Assuming $0\le \alpha_{gr}Z_T<1/2$ throughout,
the respective bound-state eigenenergies are  
\be\label{eigenen}
E_{n,j} =\frac{M}{\sqrt{1+\frac{\alpha_{gr}^2Z_T^2}{(n+\gamma_j)^2}}},\quad \gamma_j =\sqrt{j^2 -\alpha_{gr}^2Z_T^2},
\ee
where $n\ge 0$ for $j>0$, and $n>0$ for $j<0$. 
 
We next include the effects of the ion beam on the target described by $H_0$. For the conventional 3D counterpart,
electronic transitions induced by collisions of fast ( ultrarelativistic, $v\approx c$) ions with atoms have attracted much attention during the past three decades in atomic and high energy physics 
\cite{Eichler07,Eichler90,Eichler94,Eichler95}.  In particular, the 3D counterpart of the model studied below for 
the 2D graphene case was studied in Refs.~\cite{Baltz95,Baltz91,Baltz97}. We here adapt these calculations
to 2D artificial atoms, see Fig.\ref{fig1}, and discuss the corresponding physical observables. 
Following the arguments in Ref.~\cite{Baltz97}, the time-dependent potential created by the moving ion (projectile) of charge $Z_P e$ on the target electron is then given by 
\be
V(x,z,t)=-\delta(z-ct)\alpha Z_P(1-\sigma_z)\ln\left(1+\frac{x^2}{b^2}\right), \label{potential}
\ee
where $b$ is the impact parameter and $\alpha\simeq 1/137$ the standard fine structure constant.
The dynamics of the complete projectile-plus-target system  is thus described by the time-dependent Dirac equation
\be
i\frac{\partial\Psi({\bf r},t)}{\partial t}=[H_0+V({\bf r},t)]\Psi({\bf r},t), \label{DE01}
\ee
with $H_0$ in Eq.~\eqref{hamilt}. We recall that ${\bf r}=(x,z)$ is the electron position corresponding to
the target atom. 

As for the 3D counterpart \cite{Baltz97}, the  solution of Eq.~\eqref{DE01}  can be obtained by expanding $\Psi({\bf r},t)$ in terms of the complete set of eigenfunctions of $H_0$,
\be
\Psi({\bf r},t)=\sum_l a_l(t)\psi_l(\textbf{r})e^{-iE_lt}, \label{WF01}
\ee
where $a_l(t)$ are complex-valued and time-dependent expansion coefficients.  The index $l$ includes the 
quantum numbers $n$ and $j$ in Eq.~\eqref{FG}.
Substituting Eq.~\re{WF01} into Eq.~\re{DE01}, we find that 
the coefficient $a_f(t)$ for the final state $|\psi_f\rangle$ is governed by
\begin{equation}
\frac {da_{f}(t)}{dt}=-ie^{iE_f t}\langle\psi_f|V({\bf r},t)|\psi_i\rangle,\label{ampl00}
\end{equation}
where we assumed that prior to the interaction, the atomic electron was
in the state $|\psi_i\rangle$, i.e.,
\begin{equation}
\Psi_i({\bf r},t\to -\infty)= e^{-iE_it} \psi_i({\bf r}).
\label{eq:a2}
\end{equation}
Equation \eqref{ampl00} thus comes
with the initial condition
\begin{equation}
a_{f}(t\to -\infty) = \delta_{f,i}.\label{eq:a3}
\end{equation}
In order to obtain the exact solution of Eq.~\re{ampl00} with $V(x,z,t)$ in Eq.~\eqref{potential},
 it suffices to determine $\Psi({\bf r},t)$ near $z=ct$ \cite{Baltz97}. We therefore transform to light-cone coordinates,
\begin{equation}
z^-=\frac{1}{\sqrt{2}}(ct-z), \quad z^+=\frac{1}{\sqrt{2}}(ct+z).\label{eq:a4}
\end{equation}
For $t<z/c$ but very close to $t=z/c$, 
by integrating Eq.~\eqref{DE01} over $z^-$, taking into account the $\delta(z^-)$ function and the Heaviside step function $\theta(z)$ with $\frac{d}{dz}\exp\theta(z)=\delta(z)\exp\theta(z)$, we obtain
\begin{multline}
 (1-\sigma_z)\Psi_i({\bf r},t)= \\
 (1-\sigma_z)e^{-i\theta(ct-z)\alpha Z_P\ln(1+x^2/b^2)}\psi_i({\bf r})e^{-iE_it}.   
 \label{sol01}
\end{multline}
Next  we substitute Eq.~\re{sol01} into Eq.~\re{ampl00} and integrate over time. Taking into account  the initial condition \re{eq:a3}, we obtain the transition amplitude $a_{f,i}=a_f(t\to \infty)$ as
\begin{widetext}
\be
a_{f,i}=\delta_{f,i}- i\int_{-\infty}^\infty dte^{i(E_f-E_i)t} \langle\psi_f|\delta(z-ct)\alpha Z_P(1-\sigma_z)\ln\left(1+\frac{x^2}{b^2}\right)|e^{-i\theta(ct-z)\alpha Z_P\ln(1+x^2/b^2)}\psi_i\rangle. \label{ampl01}
\ee
Performing the $t$-integration,  we obtain the final result
\be
a_{f,i}=\delta_{f,i}+\langle\psi_f|(1-\sigma_z)e^{i(E_f-E_i)z/c}(e^{-i\alpha Z_P\ln(1+x^2/b^2)}-1)|\psi_i\rangle. \label{ampl02}
\ee
\end{widetext}
Equation~\re{ampl02} gives the transition amplitude of the target atom from the initial state $|\psi_i\rangle$ to the final state $|\psi_f\rangle$ induced by its interaction with the fast  ion
with charge $Z_P$.   This expression has been derived by assuming that the projectile's velocity is close to the speed of light, $v\alt c$.  We next use this expression for calculating the probabilities and cross-sections for excitation and ionization.

\emph{Probabilities and cross sections for electronic transitions.---}We now present results for the   probabilities and cross sections of target electron transitions based on the exact expression in Eq.~\eqref{ampl02}. Our calculations use the so-called survival or staying probability of the target electron in its initial state. The survival probability is the probability that the electron remains in its initial state during the interaction with the external perturbation. 
Taking into account $E_f -E_i = 0$, from  Eq.~\eqref{ampl02}, the survival probability can be written as
\be
P_i(b)=\left|1+\langle\psi_i|(1-\sigma_z)\left(e^{-i\alpha Z_P\ln(1+x^2/b^2)}-1\right)|\psi_i\rangle\right|^2. \label{prob01}
\ee
Similarly, for the transition probability from an initial state $|\psi_i\rangle$ to the
final state, $|\psi_f\rangle$ (with $f\neq i$), we obtain
\begin{eqnarray} \label{prob02}
&& P_{f,i}(b) =\\ && \left|\langle\psi_f|(1-\sigma_z)e^{i(E_f-E_i)z/c}\left(e^{-i\alpha Z_P\ln(1+x^2/b^2)}-1\right)|\psi_i\rangle\right|^2.\nonumber
\end{eqnarray}

\begin{figure}[t!]
\centering
\includegraphics[totalheight=0.24\textheight]{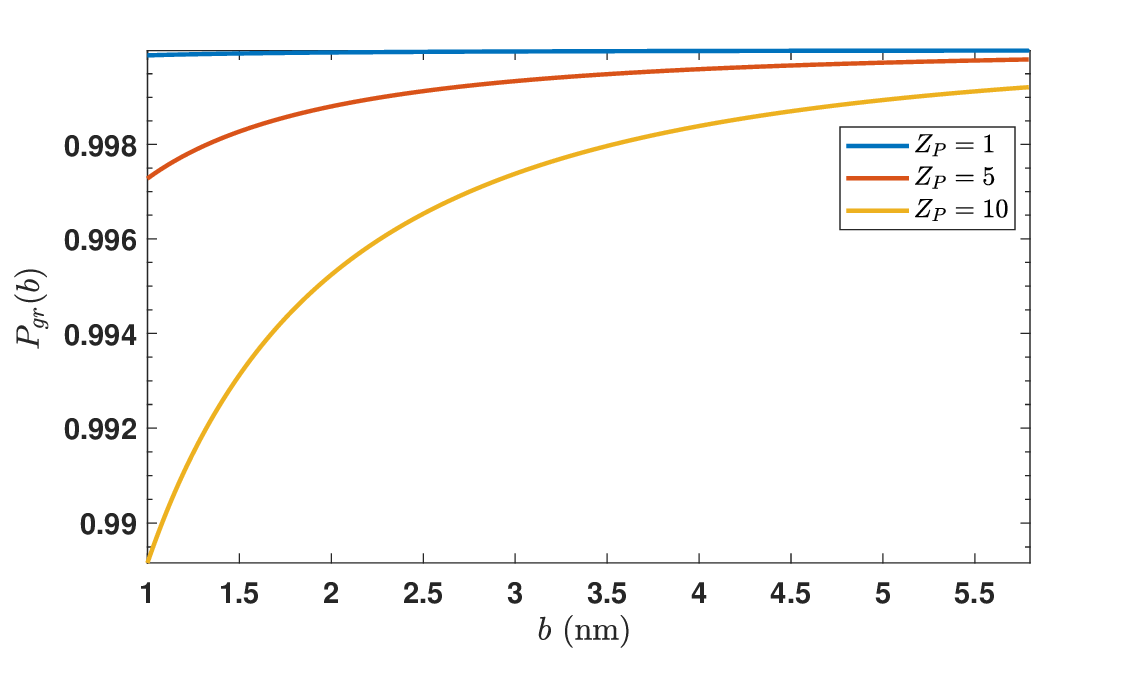}
 \caption{Survival probability in the ground state of the artificial atom, see Eq.~\eqref{prob01}, vs impact parameter $b$ 
 at different values of the projectile charge $Z_P$ in Eq.~\eqref{potential} for $\alpha_{gr}Z_T=0.4$, $M\approx0.1$ eV and $v_F=c/300$. Note the logarithmic scale for the probabilities.  } 
  \label{fig2}
\end{figure} 

\begin{figure}[t!]
\centering
\includegraphics[totalheight=0.24\textheight]{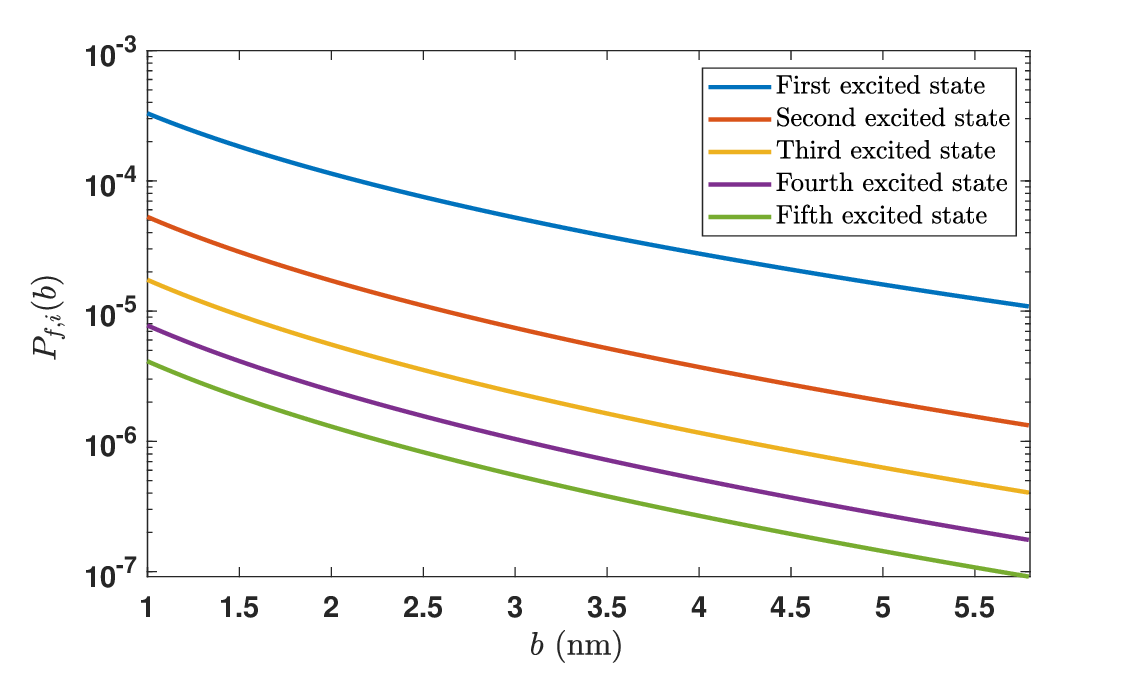}
 \caption{Excitation probability \eqref{prob02} vs impact parameter $b$ 
 from the ground state of the artificial atom to the first five excited states 
 due to the interaction with an ultrarelativistic ion with $Z_P=10$ for $Z_T\alpha_{gr}=0.4$.  The 
 other parameters are as specified in the caption to Fig.~\ref{fig2}. Note the logarithmic scale for the probabilities.
} \label{fig3}
\end{figure}

In Figure \ref{fig2}, the survival probability \eqref{prob01} of the target ground state is 
plotted as a function of the impact parameter $b$ for different values of the projectile charge $Z_P$. We observe that the survival probability grows rapidly within a small range of $b$, saturating at the maximal value $P_i(b)=1$ at large $b$.  Upon increasing the projectile charge, the staying probability is seen to decrease. Both observations are in agreement with intuitive expectations.

Next, in Figure \ref{fig3}, we show the excitation probabilities \eqref{prob02} from the ground state to the first five excited states of the artificial atom. Unlike the survival probability, a rapid decline of the probability 
with increasing impact parameter $b$ is observed.  This decrease of the excitation probability is approximately exponential in $b$, at least for not too small $b$, which can be rationalized by the finite energy differences between the ground state and the respective excited states.  In addition, the respective probabilities decrease with increasing order of the excited state.

\begin{figure}[t!]
\centering
\includegraphics[totalheight=0.24\textheight]{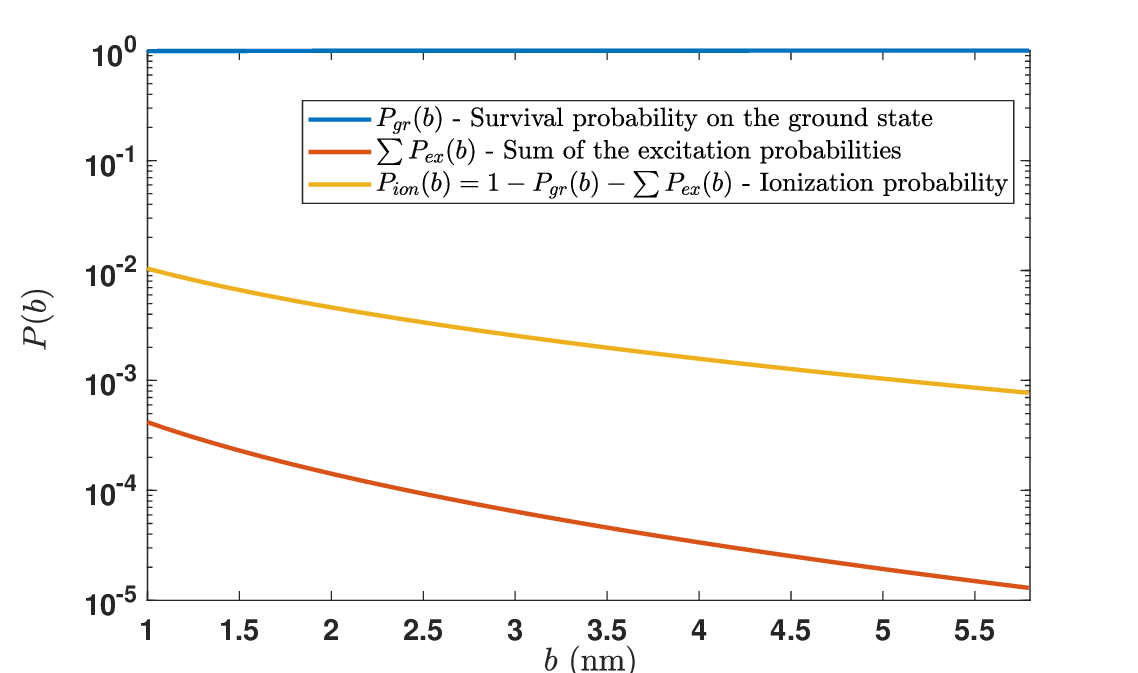}
 \caption{Survival probability \eqref{prob01}, the sum of the excitation probabilities \eqref{prob02} to excited bound states, and the total ionization probability \eqref{prob03} vs impact parameter $b$ for $Z_P=10$ and $Z_T\alpha_{gr}=0.4$. Note the logarithmic scale for the probabilities.  The other parameters are as in Fig.~\ref{fig2}.}
 \label{fig4}
\end{figure}

According to the unitarity principle,  the sum of the survival, the excitation (bound states) and the ionization (continuum states) transition probabilities should give unity. This fact implies that the total ionization probability (summed over all continuum states), $P(b)$, can be computed without explicit calculation of the respective transition  amplitudes using the 
scattering state wave functions, see also Ref.~\cite{Baltz00}. The total
ionization probability thus follows as
\be
P(b)=1-P_i(b)-\sum_{f\neq i}P_{f,i}(b), \label{prob03} 
\ee
where the sum over $f$ extends over all bound states of the artificial atom except for the ground state.
In Fig.~\ref{fig4}, we show the survival probability, the sum of the excitation probabilities, and
the total ionization probability as a function of the impact parameter $b$.  For the total excitation and ionization
probabilities, we find a decay of the probability as $b$ increases.  Importantly, the total ionization probability exceeds the total excitation probability by several orders of magnitude, especially for large $b$.

\begin{table}[t!]
\begin{center}
\vskip 0.2cm
\begin{tabular}{|c|c|}
\hline
\hspace{5mm} $Z_{P}$ \hspace{5mm} & \hspace{5mm} $\sigma$ (barn) \hspace{5mm} \\
\hline
\hspace{5mm} 1 \hspace{5mm} & $2.76 \times 10^7$ \\
\hline
\hspace{5mm} 5 \hspace{5mm} & $6.88\times 10^8$ \\
\hline
\hspace{5mm} 10 \hspace{5mm} & $2.74\times 10^9$ \\
\hline
\end{tabular}
\end{center}
\caption{Ionization cross section \eqref{cross} of the artificial atom for several values of the ion 
charge parameter $Z_P$. The same parameters as specified in Fig.~\ref{fig2} have been used. 
\label{tab1}
}
\end{table}
\begin{figure}[b]
\centering
\includegraphics[totalheight=0.17\textheight]{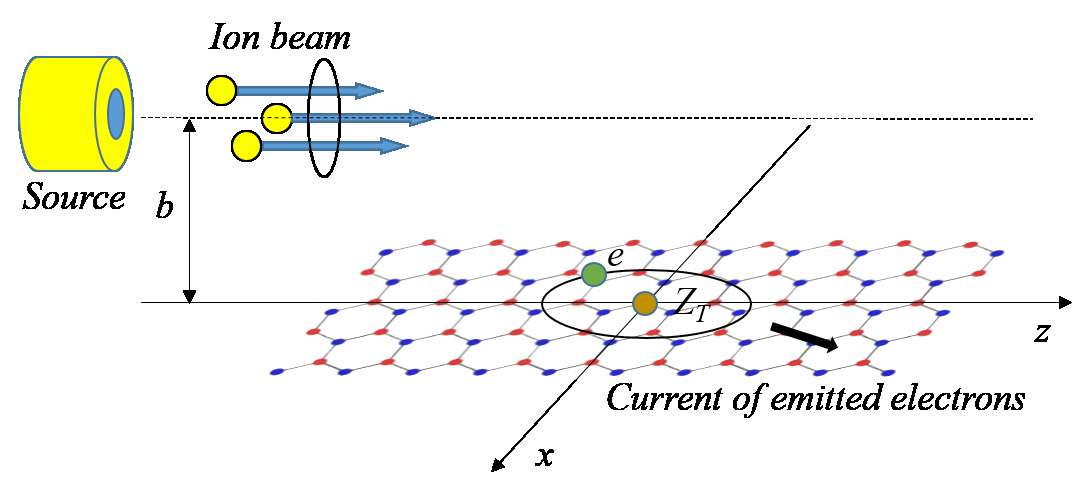}
 \caption{Sketch for a possible experimental realization of the setup studied in this work.}
 \label{fig5}
\end{figure}

From the total ionization probability $P(b)$,  the total ionization cross section follows by integrating over the impact parameter \cite{Baltz00},
\be
\sigma=2\pi \int_0^\infty db\, b  P(b). \label{cross}
\ee
In Table \ref{tab1}, we provide the resulting values of the ionization cross section for different values of the projectile charge parameter $Z_P$.  Comparing the values of the cross section to those for the 3D counterpart \cite{Baltz00}, the cross section is found to be two orders of magnitude larger.  As expected, upon increasing $Z_P$, the cross section also increases substantially.

\emph{Conclusions.---}In this work, we have considered a low-energy theory for a Coulomb impurity in a graphene monolayer interacting with a fast (ultrarelativistic) ion moving parallel to the graphene sheet at distance $b$. 
The exact solution of the time-dependent 2D Dirac equation describing the interaction of the target artificial atom (which is created by the Coulomb impurity) with the ultrarelativistic charged projectile is obtained. Using this solution, the electronic transition probabilities to excited bound states, the survival probability in the ground state, and the total ionization probability have been analyzed as function of the impact parameter $b$. In addition, we have computed the cross section which is found to be  two orders larger than its 3D counterpart. Probing collision-induced electronic transitions in the planar atoms formed by Coulomb impurities in graphene formed offers interesting perspectives for future experiments.    

Let us also mention that the above approach can be applied for the calculation of arbitrary electronic transitions, including particle-hole pair creation processes and the angular distribution of differential cross sections. 
An interesting extenstion of our study is to investigate the supercritical case $\alpha_{gr} Z_T>1/2$, see, e.g., Refs.~\cite{Wang,Pereira,Denis,Asorey20,Kuleshov17}. In this case as the target state wave functions one can use those obtained in the Refs.~\cite{Pereira,Denis,Asorey20}. The final state wave function (for excitation) one can use either discrete state eigenvalues of the Dirac operator given by \re{hamilt}, or plane wave states considered in \cite{Novikov}. All these can be done within the above approach.
Experimentally, the above model could be realized by passing a fast charged ion (e.g, proton) beam over the graphene layer, as schematically presented in Fig.~\ref{fig5}.

\emph{Acknowledgements.---}We acknowledge funding by the Grant REP-05032022/235 ("Ultrafast phenomena and vacuum effects in relativistic artificial atoms created in graphene"), funded under the MUNIS Project, supported by the World Bank and the Government of the Republic of Uzbekistan. 
The work of DM is supported by the grant of the Ministry for Higher Education, Research and Educations (FZ-5821512021).


\end{document}